# Semiconductor-to-Metal Transition in the Bulk of $WSe_2$ upon Potassium Intercalation

Mushtaq Ahmad[1,2], Eric Müller[1], Carsten Habenicht[1], Roman Schuster[1], Martin Knupfer[1], and Bernd Büchner[1]

[1]IFW Dresden, D-01069 Dresden, Germany

[2]Department of Physics, COMSATS Institute of Information Technology, Park Road, 45550, Islamabad, Pakistan

## Abstract

We present electron energy-loss spectroscopic measurements of potassium (K) intercalated tungsten diselenide ($WSe_2$). After exposure of pristine $WSe_2$ films to potassium, we observe a charge carrier plasmon excitation at about 0.97 eV, which indicates a semiconductor to metal transition. Our data signal the formation of one particular doped K-$WSe_2$ phase. A Kramers-Kronig analysis (KKA) allows the determination of the dielectric function and to estimate the composition of about $K_{0.6}WSe_2$. Momentum dependent measurements reveal a substantial plasmon dispersion to higher energies. .

## Introduction

In the last few years, $WSe_2$ belonging to the family of two-dimensional (2D) transition metal dichalcogenides (TMDCs) has attracted significant attention both theoretically and experimentally, because of its unique electronic structure, high carrier mobilities, and quantum relativistic phenomena [1]. In particular, the possibility to mechanically exfoliate atomically thin sheets of $WSe_2$ has caused particular interest [2]. Such 2D materials possess various fascinating anisotropic properties [3, 4], tunable bandgaps and strong light-matter coupling [5] which renders them one of the important materials to be used for future electrical and optical devices. In addition to that, being a 2D material and having a suitable bandgap, which is mandatory for many applications, fills the gap left by graphene. $WSe_2$ in bulk form is a *p*-type material with an indirect band gap of ~1.2 eV, however, its monolayer structure yields a direct band gap of ~1.65 eV [6]. The direct band gap exists around the *K*-points of the Brillouin zone between the spin-orbit split valence band and the doubly degenerate conduction band, the indirect band gap arises between the valence band maximum at the $\Gamma$ point and the conduction band minimum at mid of $\Gamma$ to $K$ [7].

The intercalation of layered structures with alkali metals often has opened a route to new materials with interesting physical properties. Specifically due to its small electron affinity, potassium is considered as a strong electron donor to most surfaces and in the bulk. A prominent example is the potassium intercalation of graphite and the resulting stable phases which e.g. become superconducting [8]. Also dichalcogenide materials have been intercalated with potassium, resulting in the formation of metallic $K_{0.4}MoS_2$ and $K_{0.5}WS_2$ compounds, whereas the former also supports a superconducting ground state [9, 10]. More recently, *p*-doping with a molecular compound [11] and degenerate *n*-doping of $WSe_2$ [12] in field effect transistors was reported. In addition, surface electron doping of $WSe_2$ by potassium has resulted in the observation of a negative electronic compressibility in the doped surface [13].

Building on this concept, we have investigated the electronic excitation spectrum of bulk 2H-$WSe_2$ intercalated with potassium using electron energy-loss spectroscopy (EELS) in transmission. We observe the appearance of a new excitation at around 1 eV, which is indicative for the formation of $K_{0.6}WSe_2$ and for a semiconductor to metal transition upon doping. A Kramers-Kronig analysis provides the dielectric function of this material.

**Experimental details**

Firstly, high quality thin films of a thickness of about 100 nm were prepared via mechanical exfoliation from single crystals, which were purchased from "2D Semiconductors" [14]. Thereafter, the films were mounted onto standard electron microscopy grids and subsequently transferred in suitable sample holders, and then finally loaded into the 172 keV spectrometer for EELS measurements [15]. The purpose built EEL spectrometer, which is capable of precise measurement of the materials' dynamic response, is equipped with a helium flow cryostat, and all measurements were carried out at 20 K to have minimum thermal broadening. EELS in transmission is a bulk sensitive scattering technique that measures the electronic excitations of the material under investigation and provides the loss function $L(q,\omega) = \text{Im}(-1/\varepsilon(q,\omega))$, where $\varepsilon(q,\omega)$ is the momentum ($q$) and energy ($\omega$) dependent dielectric function [16]. In this work, the energy ($\Delta E$) and momentum ($\Delta q$) resolution of the instrument were chosen to be 82 meV and 0.04 Å$^{-1}$ respectively. Further details about the spectrometer and its characteristics are reported elsewhere [15, 17]. Recently, this technique has also been employed to study topical quasi-2D materials such as $MoS_2$ [18], black phosphors [19], as well as e.g. organic materials [20-23].

Potassium intercalation of $WSe_2$ films was carried out via thermal evaporation of potassium under ultra-high vacuum conditions (base pressure below $10^{-10}$ mbar), where the films were mounted above the applied K dispensers (SAES Getters, S.P.A, Italy). A current of 6.4 A was applied to the K dispenser to initiate the intercalation process. The intercalation was performed in several steps, each lasted about of 30 - 45 seconds. In order to achieve a homogenous distribution of potassium, the films were annealed at 80 °C for 1 hour in an oven within the vacuum chamber.

**Results and discussions**

Once we had verified the pure hexagonal phase of $WSe_2$ by measuring the electron diffraction profile, we measured the loss function. Figure 1 shows the energy-loss spectra of pure and K doped $WSe_2$ in a wide energy range up to 70 eV along the ΓM direction of the Brillouin zone. The momentum transfer is 0.1 Å$^{-1}$, which corresponds to the optical limit [15].

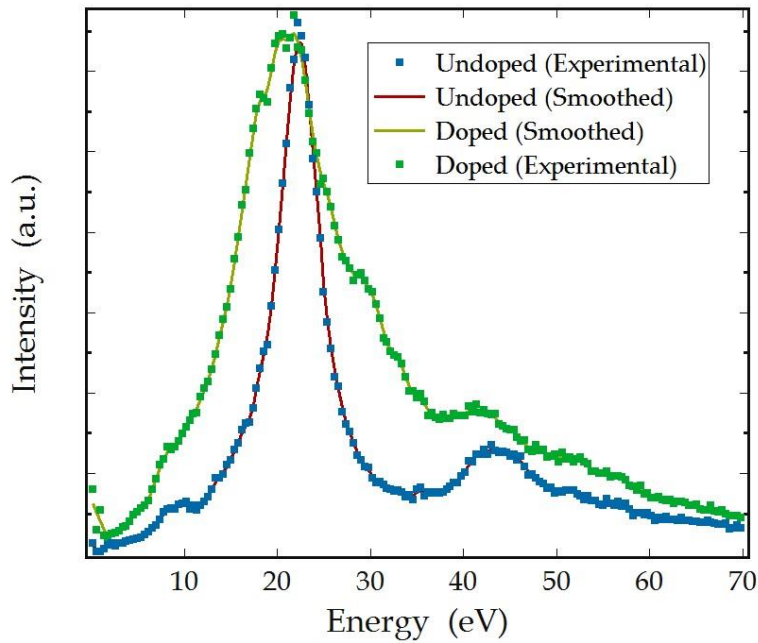

Figure 1. Electron energy-loss spectra of pure and intercalated $WSe_2$ at 20 K taken with a momentum transfer of 0.1 Å$^{-1}$, which corresponds to the optical limit. The measured data are shown as dots, a smoothed line is depicted as a guide to the eye.

The loss function of pristine WSe$_2$ is dominated by a strong feature at about 22 eV which can be associated with the volume plasmon, a collective excitation of all valence electrons. At around 44 eV, a broad feature is visible which is caused by a superposition of multiple scattering and shallow core level excitations from the W 5p levels. At low energies (see Fig. 2), the loss function of WSe$_2$ is characterized by an excitation at 1.8 eV followed by a second feature at about 2.3 eV. These two excitations are due to excitons related to transitions at the K point of the Brillouin zone, whereas their energy difference represents the spin-orbit splitting of the valence band edge [24-26].

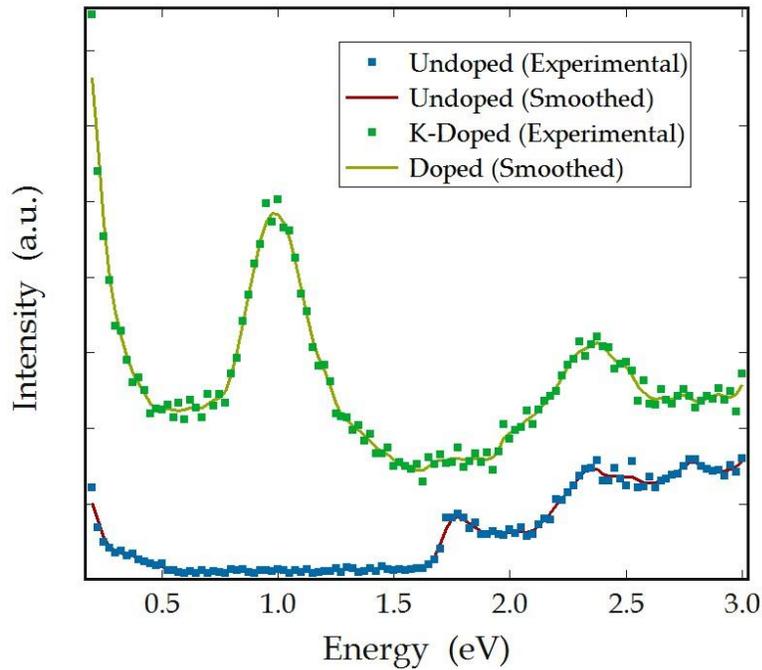

Figure 2. Electron energy-loss spectra of pure and intercalated WSe$_2$ in the lower range. The momentum transfer is 0.1 Å$^{-1}$. The measured data are shown as dots, a smoothed line is depicted as a guide to the eye.

Potassium intercalation and the concomitant charge transfer to WSe$_2$ results in substantial changes in the loss function. The broad structure around 21 eV again arises from the volume plasmon of K intercalated WSe$_2$, and in addition excitations from K 3$p$ levels contribute at energies around 18 eV. Going to the low energy window in Fig. 2, a clear potassium induced excitation feature becomes visible at ~ 0.97 eV. In agreement with the occurrence of an insulator to metal transition in related materials such as K intercalated $WS_2$ [10] or $MoS_2$ [9] we assign this new excitation to the charge carrier plasmon in the case of doped, metallic $WSe_2$.

This appearance of a plasmon excitation for potassium doped $WSe_2$ at 0.97 eV is present in the measured spectra after each step of K addition, whereas the energy position is independent of the doping or intercalation level. This strongly indicates the formation of a well-defined intercalated K-$WSe_2$ phase, which is formed in our samples at expense of undoped $WSe_2$ until the entire sample has been transformed to this K-$WSe_2$ phase. This behavior parallels that observed for other potassium intercalated materials, where phases with defined compositions are formed [27, 28]. Interestingly, in the case of K intercalation of the metallic dichalcogenides $NbSe_2$ and $TaSe_2$, exposure to potassium lead to homogeneous K intercalated compounds at arbitrary intercalation level [29] in contrast to the formation of particular phases as observed for $WSe_2$ here. Furthermore, the plasmon energy of about 1 eV as observed for K intercalated $WSe_2$ in this work is very similar to the plasmon energies in metallic, undoped dichalcogenides, e.g. $NbSe_2$ or $TaSe_2$ [30]. In the latter the charge carrier plasmon represents the collective excitations of a half filled 4d- or 5d-derived conduction band.

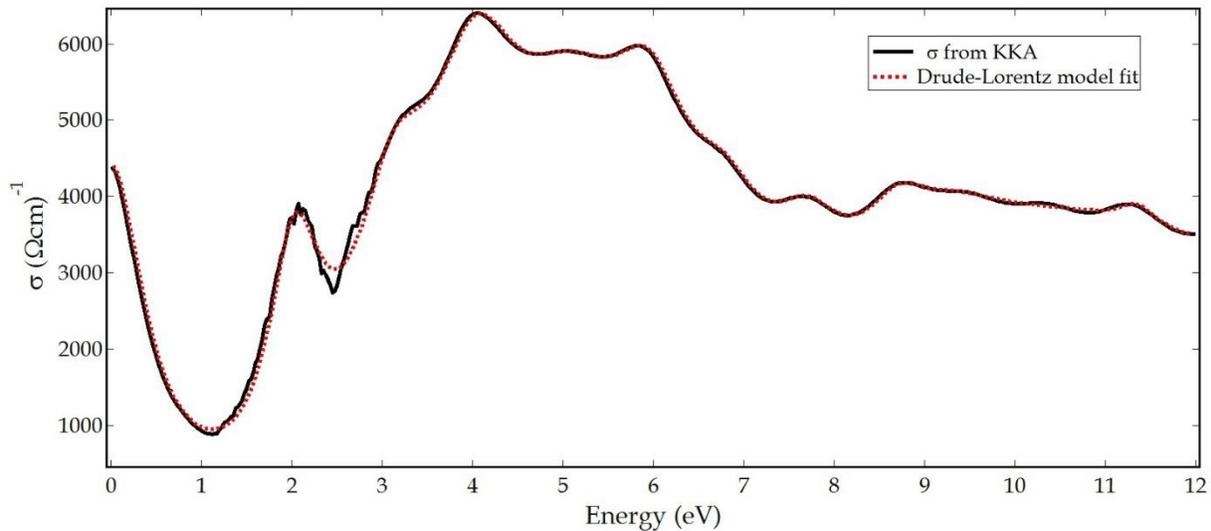

Figure 3. The optical conductivity (straight line) deduced from a KKA and its very well fitting (dashed lines) with Drude-Lorentz model.

In order to achieve more insight into the electronic properties of our K-$WSe_2$ compound, we have carried out a Kramers-Kronig analysis (KKA) of the measured loss function under the assumption of a metallic ground state. For that, the energy loss function measured in the range 0 – 70 eV is corrected for the elastic line contribution and multiple scattering effects according to the procedure outlined in previous reports [15, 26]. We additionally have carried out a KKA under the assumption that our K-WSe$_2$ compound is characterized by a small band gap (200 meV), which we would not be able to see in the data due to the elastic line contribution. This then resulted in unphysically large values (> 50) of the real part of the dielectric function at small energies, which provides clear evidence that our data stem from a metallic material.

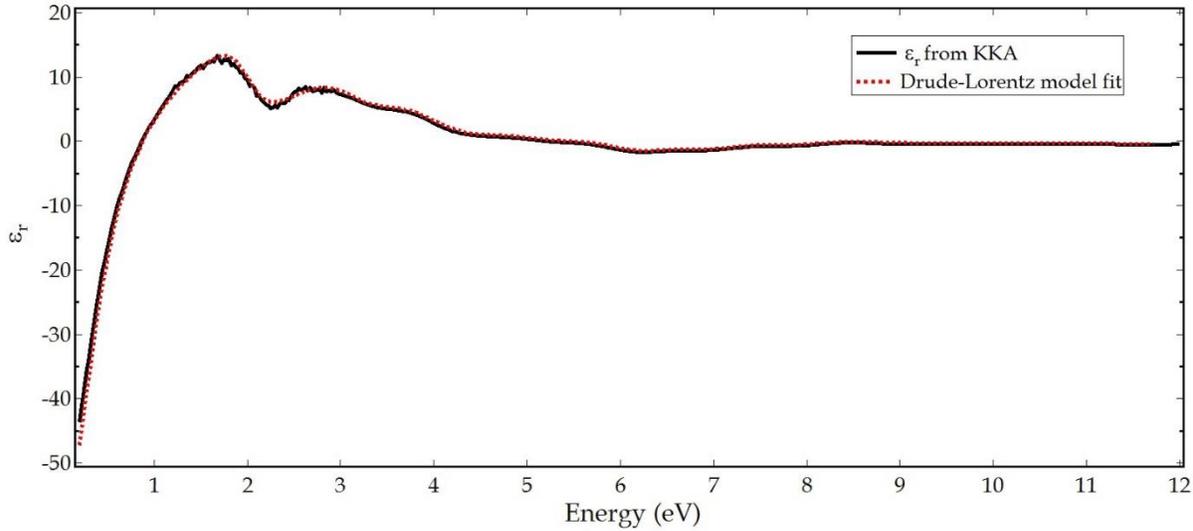

Figure 4. The real part of dielectric function (straight line) deduced from a KKA and its fitting (dashed lines) with Drude-Lorentz model.

In order to separate the charge carrier plasmon from the single particle excitations, we have fitted the resulting optical conductivity $\sigma = \omega \varepsilon_2$ and the real part of the dielectric function ($\varepsilon_r$) using a Drude-Lorentz model between 0 and 12 eV. The Drude-Lorentz model can be described as;

$$\varepsilon(\omega) = \varepsilon_\infty + \sum_n \frac{\omega^2_{pn}}{\omega^2_{0n} - \omega^2 - i\gamma_n \omega}$$

Here, $\varepsilon_\infty$, $\omega$, $\omega_p$, $\omega_0$, $\gamma$, and $n$ denote the background dielectric constant, energy, plasma frequency (oscillator strength), energy position, spectral width and each (numbered by n) oscillator respectively. Figure 3 shows the optical conductivity as obtained by the KKA and the results of our fitting procedure. It becomes clear that our Drude-Lorentz approach provides a very good description of the optical conductivity. The conductivity at energies up to about 1 eV represents the Drude part related to the charge carrier plasmon, whereas beyond this energy range the Lorentz oscillators explain the conductivity arising from single particle excitations. Figure 4 shows the real part of dielectric function $\varepsilon_r$ calculated from the KKA and the corresponding Drude-Lorentz fit using the same parameters as for the optical conductivity. In addition, we allowed for the variation of a background dielectric function ($\varepsilon_\infty$), which contributes to the real part only and a value of 1.48 is obtained. This background dielectric constant describes the screening aptitude of interband transitions above the considered energy range [31]. All fitted parameters are summarized in Table 1.

Table 1. Oscillator fit parameters of the optical conductivity and the real part of the dielectric function of potassium doped WSe$_2$. In addition, a background dielectric constant of 1.48 was obtained. All parameters are given in eV.

| Oscillator | $\omega_{pn}$ | $\omega_{0n}$ | $\gamma_n$ |
|---|---|---|---|
| Drude | 3.82 | 0.00 | 0.44 |
| Lorentz 1 | 3.90 | 2.04 | 0.72 |
| Lorentz 2 | 4.82 | 3.14 | 1.14 |
| Lorentz 3 | 6.02 | 4.05 | 1.33 |
| Lorentz 4 | 6.52 | 5.13 | 1.88 |
| Lorentz 5 | 4.05 | 5.97 | 1.08 |
| Lorentz 6 | 2.76 | 6.79 | 1.03 |
| Lorentz 7 | 2.09 | 7.70 | 0.86 |
| Lorentz 8 | 1.82 | 8.72 | 0.79 |
| Lorentz 9 | 1.98 | 9.33 | 1.41 |
| Lorentz 10 | 14.7 | 10.84 | 8.75 |
| Lorentz 11 | 0.92 | 11.35 | 0.48 |

The results of our fitting procedure now allow for further insight into the properties of K doped WSe$_2$. Using the fitted value of the charge carrier (Drude) plasmon energy ($\omega_p$), and taking into account the elementary charge $e$, the electron mass $m$ and free space permittivity $\varepsilon_0$ in the relation $\omega^2_p = ne^2/\varepsilon_0 m$, we calculated the density $n$ of conduction electrons per unit cell of our doped WSe$_2$. In order to obtain the number of electrons per unit cell, we have considered a lattice expansion in the direction perpendicular to the WSe$_2$ layers of about 30 %, as it is observed for other potassium intercalated dichalcogenides [9, 29]. This then results in about 1.2 electrons/unit cell and would correspond to a composition of about K$_{0.6}$WSe$_2$ assuming full charge transfer from K to WSe$_2$. Previously, close relatives to WSe$_2$, MoS$_2$ and WS$_2$, which are structurally and electronically very similar, have also been intercalated, and compositions of K$_{0.4}$MoS$_2$ [9] and K$_{0.5}$WS$_2$ [10] have been reported in very good correspondence to our results. We emphasize that the evolution of our data as a function of potassium addition strongly indicates that this is the only K doped phase of WSe$_2$ which is formed.

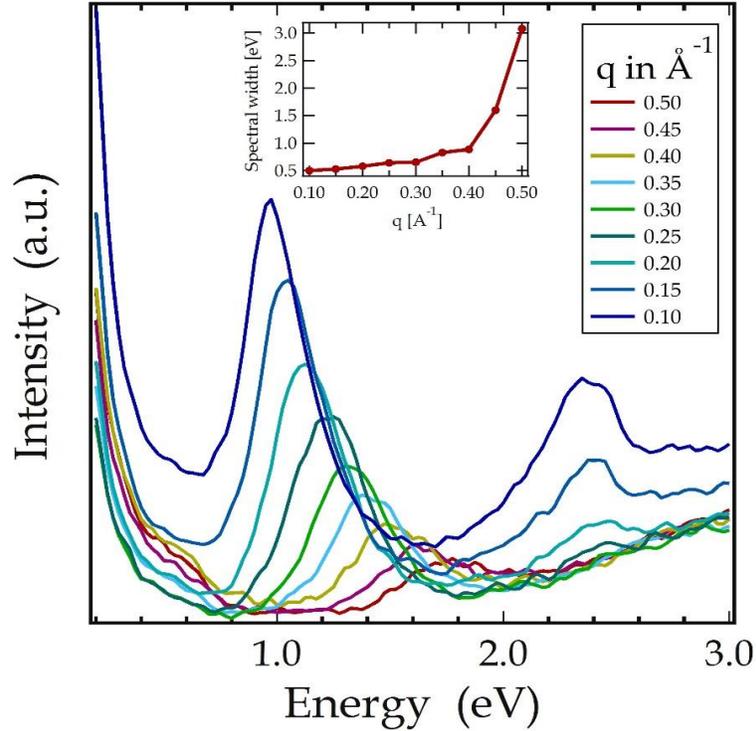

Figure 5. Plasmon dispersion of K intercalated WSe$_2$ along ΓM direction of Brillouin zone at 20 K. We observed an equivalent plasmon dispersion along the ΓK direction as well as at room temperature.

Finally, we turn our attention to the momentum dependence of the charge carrier plasmon, i.e. its dispersion, in potassium doped $WSe_2$. In Fig. 5 we present the evolution of the loss function in the corresponding energy range upon increasing momentum transfer. The data clearly show an upward energy shift of the charge carrier plasmon with increasing momentum transfer. We have also determined the spectral width of the plasmon which is depicted in the inset of Fig. 5. These data demonstrate that the plasmon width remains rather constant up to a momentum transfer of about 0.4 Å$^{-1}$, before it significantly broadens. This indicates that above 0.4 Å$^{-1}$ the plasmon life time becomes substantially reduced and we assign this behavior to the fact that the plasmon enters the continuum of single particle excitations.

In addition, Fig. 5 also shows that the second spectral feature around 2.4 eV, which must be related to single particle excitations decreases significantly with increasing momentum transfer and has disappeared above about 0.25 Å$^{-1}$. Such a strong momentum dependent intensity variation has been observed for materials with quite localized excitations such as molecular solids [32, 33], but it is rather unusual for materials with strong covalent bonds and a well defined electronic band structure with a sizable band width.

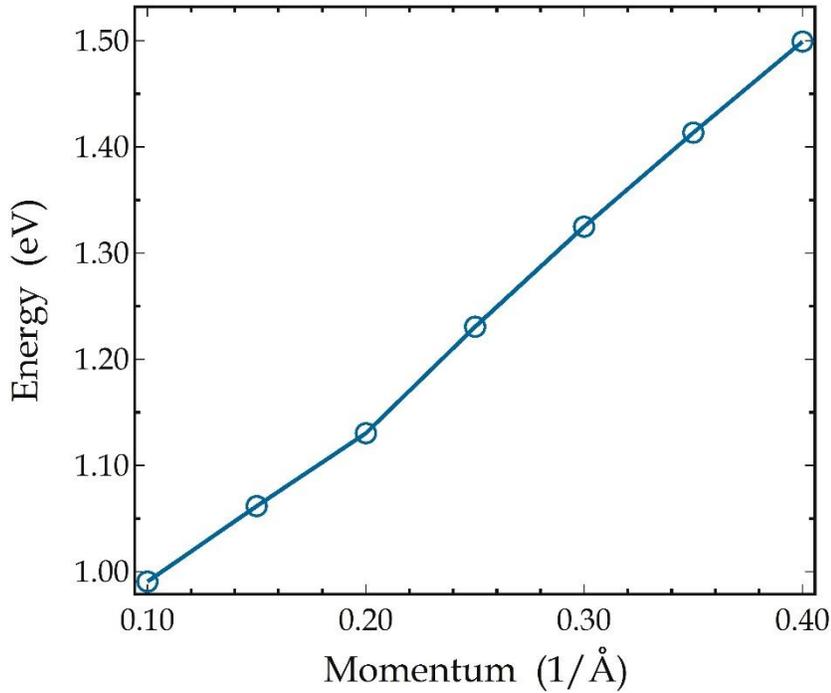

Figure 6. Summary of plasmon energy positions (plasmon dispersion) as a function of momentum along the ΓM direction of the hexagonal Brillouin zone.

Figure 6 summarizes the energy position (plasmon dispersion) along the ΓM direction of the hexagonal Brillouin zone at 20 K for momenta up to 0.4 Å$^{-1}$, where the plasmon is a sharp well defined spectral feature. We observed an equivalent plasmon dispersion along the ΓK direction as well as at room temperature. It becomes clear that K doped WSe$_2$ harbors charge carrier plasmon excitations with a clear positive dispersion. Moreover, our data suggest a quasi-linear increase of the plasmon energy upon increasing momentum. While the plasmon dispersion would be expected to show a quadratic increase with momentum for a homogeneous electron gas, such a quasi-linear increase has also been observed quite frequently and in some cases it could be related to the band structure of the respective materials [34-36].

Intriguingly, the plasmon dispersion of the undoped, metallic dichalcogenides NbSe$_2$, TaSe$_2$ and TaS$_2$ has been found to be negative [30, 37], in stark contrast to the observations here and the expectations for homogeneous electron gas systems [38]. For these compounds, it has been argued that this is a consequence of the particular band structure giving rise to band-band transitions that impact the plasmon and in particular its momentum dependence strongly [39, 40]. Upon potassium or sodium intercalation, the plasmon dispersion in NbSe$_2$, TaSe$_2$ and TaS$_2$ becomes positive but never reaches a slope as large as observed here. In potassium doped WSe$_2$ the 5d-derived conduction bands are filled with more than two electrons per formula unit, significantly more than the single electron per formula unit in the conduction bands of NbSe$_2$, TaSe$_2$ and TaS$_2$, and the band structure close to the Fermi level thus will also differ accordingly, which provides a natural basis to explain the different plasmon dispersions. Moreover, the strong intensity decrease of the excitation around 2.4 eV might support a strong, positive plasmon dispersion since it signals reduced plasmon screening at higher momenta.

**Summary**

We have carried out investigations of the electronic properties of WSe$_2$ when doped via the intercalation with potassium using electron energy-loss spectroscopy (EELS) in transmission. Our data reveal the appearance of a doping induced charge carrier plasmon at about 1 eV which suggests a semiconductor to metal transition as a consequence of potassium addition. Furthermore, the evolution of the electronic excitations provides evidence for the formation of a single intercalated phase with a composition of about K$_{0.6}$WSe$_2$. The plasmon dispersion is found to be quasi-linear and positive.

## Acknowledgements

Mushtaq Ahmad would like to acknowledge the financial support of German Academic Exchange Service (DAAD) under the Leibniz-DAAD post-doctoral fellowship program. Funding through the IFW excellence program is also acknowledged. We are thankful to M. Naumann, R. Hübel, and S. Leger for their technical support.